\documentclass[preprint]{elsarticle}
\usepackage[T1]{fontenc}
\usepackage[utf8]{inputenc}
\usepackage{amsmath,amssymb,amsthm}
\usepackage[english]{babel}
\usepackage{graphicx}
\usepackage{xcolor}
\usepackage{import}
\usepackage{bm}
\usepackage{xifthen}
\usepackage{subfig}
\usepackage{graphicx}
\usepackage{orcidlink} 
\usepackage{natbib}
\usepackage{hyperref}
\usepackage[a4paper, margin=2cm]{geometry}

\newcommand{\mrm}[1]{\mathrm{#1}}

\newcommand{\todo}[1][]{
	\ifthenelse{\isempty{#1}}%
	{{\color{magenta}\textbf{!!~}}}%
	{{\color{magenta}\textbf{#1}}}%
}

\let\bm\boldsymbol

\newcommand{\T}{^{\mathsf{T}}}

\newcommand{\at}{{\,\vert\,}}

\title{Vertical Vibration Reduction of Maglev Vehicles using Nonlinear MPC}

\author{Mario Hermle\orcidlink{0009-0004-7937-1943}}
\author{Arnim Kargl\orcidlink{0009-0003-0824-339X}}
\author{Peter Eberhard\orcidlink{0000-0003-1809-4407}}
\address{Institute of Engineering and Computational Mechanics (ITM), University of Stuttgart, Germany\\
\textit{\{mario.hermle, arnim.kargl, peter.eberhard\}@itm.uni-stuttgart.de}}

\makeatletter
\def\ps@pprintTitle{%
  \let\@oddhead\@empty
  \let\@evenhead\@empty
  \def\@oddfoot{%
    \footnotesize\itshape
    Preprint submitted to PAMM\hfill
    June 24, 2025%
  }%
  \let\@evenfoot\@oddfoot
}
\makeatother
 
\begin{document}

\begin{frontmatter}
     \begin{abstract}
          This work presents a novel Nonlinear Model Predictive Control (NMPC) strategy for high-speed Maglev vehicles that explicitly incorporates mechanical suspension dynamics into the control model. 
          Unlike conventional approaches, which often neglect the interaction between levitation magnet and car body motion, the proposed method enables predictive vibration mitigation by modeling both electromagnetic forces and suspension behavior. 
          This integrated approach significantly improves passenger comfort and ride quality by reducing vertical oscillations caused by track irregularities. 
          Moreover, it allows for a more effective tuning of the trade-off between precise air gap tracking and ride comfort. 
          Simulations based on a detailed multibody model of the Transrapid demonstrate that the method outperforms existing controllers in vibration suppression, making it a promising solution for future high-speed Maglev applications.
     \end{abstract}
\end{frontmatter}

\section{Introduction}
Electromagnetic suspension (EMS) Maglev systems, such as the Transrapid, rely on active control to maintain a stable levitation air gap between the vehicle and the guideway. 
This is essential for system stability, as the open-loop dynamics are inherently unstable. 
At the same time, the controller must ensure ride comfort by limiting accelerations transmitted to the vehicle, especially to the car body.
In this context, low frequency car body movements up to frequencies of $5\,\mrm{Hz}$ drastically reduce ride comfort for the passengers~\cite{ConollyEtAl15,LiuEtAl20}.
While there exist various active control strategies for vibration suppression of high-speed trains, they usually require additional actuators and sensors for the suspension, which increases the system complexity and cost~\cite{ZhangKordestaniShadabfar23}.
In contrast, the approach presented in this study utilizes the existing hardware to achieve a reduction in vertical vibrations solely through adaptation of the levitation control algorithm.
The central challenge lies in balancing two objectives, namely precise tracking of the nominal air gap and suppression of vertical vibrations caused by guideway irregularities. 
These irregularities, resulting from structural tolerances and elastic deflection of the guideway, are unknown to the controller and must be compensated in real time.
This study investigates the use of Nonlinear Model Predictive Control (NMPC) to address this challenge. 
NMPC offers a structured framework for optimizing multiple objectives subject to physical constraints and nonlinear system dynamics. 
In particular, and this is the novel contribution of this paper, it allows for incorporating both magnet and car body dynamics, including the suspension, into a controller model, which enables a direct influence on ride comfort.
The following sections describe the modeling approach, control design, and simulation results with a detailed multibody vehicle model that demonstrate the effectiveness of the proposed NMPC scheme in achieving stability and increased passenger comfort.

\section{Air Gap Control of Maglev Systems}\label{sec:air_gap_control}
\begin{figure}
     \includegraphics[width=\linewidth]{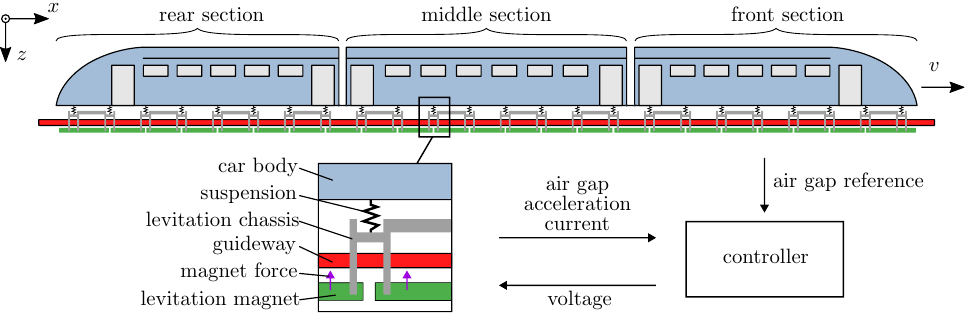}
     \centering
     \caption{Schematic of the Transrapid vehicle in the $x$-$y$-plane.}
     \label{fig:2D_model_schematic}
 \end{figure}
The Maglev vehicle model considered in this study is based on the Transrapid. 
A typical vehicle configuration consists of three sections as illustrated in Figure~\ref{fig:2D_model_schematic}.
The car body is connected to the levitation and guidance magnets through the levitation chassis.
While the levitation magnets attract the vehicle towards the guideway from below, the guidance magnets keep it aligned within the guideway. 
A soft suspension between car body and levitation chassis ensures ride comfort by dampening the high frequency vibrations of the magnets.
For a more detailed description of the Transrapid vehicle, see~\cite{DellnitzEtAl12}.
For this study, the focus is placed on the levitation aspect.
Due to their design principle, electromagnetic suspension type Maglev vehicles such as the Transrapid require active control of the magnetic forces to achieve stable levitation, as the open-loop system is inherently unstable.
However, this design allows for levitation from standstill~\cite{HermleKarglEberhard25} and minimizes issues related to magnetic flux leakage, which are more pronounced in electrodynamic suspension (EDS) systems~\cite{LeeKimLee06}. 

The levitation control algorithm consequently pursues two primary objectives. 
Firstly, the controller must ensure the stabilization of the vehicle at the nominal air gap, maintaining this equilibrium despite the presence of external disturbances. 
The nominal air gap is defined as the desired distance between the electromagnets and the guideway. 
Secondly, the controller aims to minimize accelerations experienced at both the magnet and car body levels to enhance passenger comfort.
These two objectives represent conflicting control goals that require careful balancing.

The guideway consists of individual girders with fixed length. 
However, due to structural bending and construction imperfections, the guideway surface is not perfectly smooth but exhibits periodic deviations. 
These irregularities directly influence the measured air gap for the controller and, consequently, affect the control error. 
However, the guideway profile is not known to the controller, which means it can only react to such disturbances rather than anticipate them.
An ideal tracking controller would perfectly follow the air gap reference, thereby maintaining zero control error. 
This would lead to high accelerations of both the magnet and the car body, adversely affecting ride comfort. 
In contrast, an ideal acceleration-minimizing controller would smooth out the effects of the guideway irregularities completely.
However, this may result in excessively large variations in the air gap, potentially compromising system stability. 
Therefore, a trade-off is required, yielding a controller that effectively stabilizes the system while also minimizing accelerations to enhance ride comfort.

In this study, the problem is addressed using Nonlinear Model Predictive Control (NMPC).
NMPC can explicitly account for the system nonlinearities arising in the electromagnet dynamics.
Furthermore, it is able to incorporate constraints on system states and control inputs, which is necessary in this case as the applied voltage to the electromagnets is limited.
The predictive capability makes NMPC particularly well-suited for balancing the aforementioned competing objectives through tuning of a cost function.
While a soft suspension is able to decouple the motion of the levitation chassis and the car body for high frequencies, the damping of low frequencies is limited.
This motivates the development of an algorithm, which is able to dampen these low frequency vibrations.

\section{Modeling}\label{sec:modeling}
\begin{figure}
     \centering
     \subfloat[Single mass model\label{fig:single_mass_model}]{\includegraphics{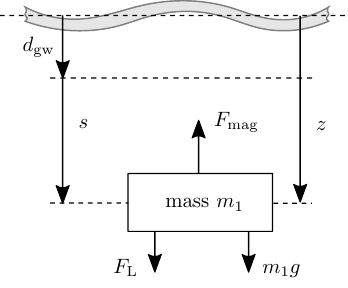}}
     \hspace{1cm}
     \subfloat[Two-mass model\label{fig:two_mass_model}]{\includegraphics{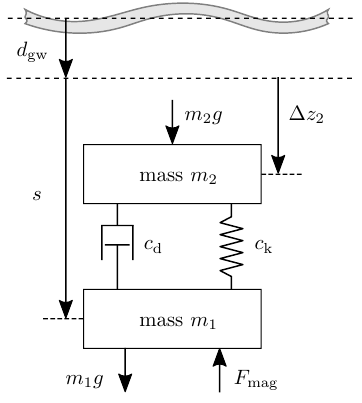}}
     \caption{The single mass model contains one unsprung mass representing one levitation half magnet. The two-mass model contains one additional sprung mass representing part of the car body's mass.}
     \label{fig:model_comparison}
 \end{figure}

The modeling of the control subsystem is divided into two main components. 
First, a mechanical model of the system is necessary to capture the dynamic interaction between the levitation magnets and the vehicle structure. 
This model serves as the foundation for predicting the vertical motion of the system.
Following the principle of the magnetic wheel~\cite{GottzeinMeisingerMiller80}, the single mass model as depicted in Figure~\ref{fig:single_mass_model} is derived.
There, the mass $m_1$ accounts for one half of a levitation magnet and one quarter of a levitation chassis. 
It is influenced by gravity and pulled towards the guideway from below by the magnet force~$F_\mrm{mag}$.
The car body and the weight of the passengers is modeled with the static load force~$F_\mrm{L}$.
The absolute position of mass $m_1$ in the inertial frame is $z_1 = d_\mrm{gw}+s$, where $s$ is the air gap between magnet and guideway and~$d_\mrm{gw}$ the guideway deflection.
This single mass model has been successfully implemented and deployed in previous studies~\cite{SchmidEberhardDignath2019,KarglEtAl25}, however, with limited capabilities on influencing the car body dynamics.
While the levitation magnets’ vertical acceleration~$\ddot{z}_1$ can be slightly suppressed with this modeling approach, almost no improvement in ride comfort, being determined by the car body’s movement, may be achieved.
Since the single mass model only takes into account the motion of the levitation magnet, the car body movement cannot directly be influenced by the controller.

Consequentially, the up to now not used two-mass model as illustrated in Figure~\ref{fig:two_mass_model}, is introduced.
The unsprung mass is extended with a mass $m_2$, representing part of the car body mass.
The two masses are connected via linear spring and damper elements, thereby approximating the vehicle suspension.
This apparently simple approach of modeling the vehicle's suspension for MPC design is also common in active suspension control of automobiles~\cite{GoehrleEtAl22}.
The spring and damping constants~$c_\mrm{k}$ and~$c_\mrm{d}$ are tuned to match the first eigenfrequency and damping ratio of the real vehicle's suspension.
While the real suspension dynamics is nonlinear, this simplification provides a sufficient level of detail to reflect the low frequent car body movements while remaining computationally tractable within the NMPC framework.
The equations of motion in vertical direction for the two-mass model are
\begin{equation}\label{eq:two_mass_mechanics}
     \begin{aligned}
         m_1 \ddot{z}_1 &= m_1 g - c_{\mrm{k}} (z_1 - z_2) - c_{\mrm{d}} (\dot{z}_1 - \dot{z}_2) - F_{\mrm{mag}}, \\
         m_2 \ddot{z}_2 &= m_2 g + c_{\mrm{k}} (z_1 - z_2) + c_{\mrm{d}} (\dot{z}_1 - \dot{z}_2)
     \end{aligned}
 \end{equation}
 with the absolute positions $z_1 = s+d_\mrm{gw} = s_\mrm{nom} + \Delta s + d_\mrm{gw}$ and $z_2 = \Delta z_2+d_\mrm{gw}$ of $m_1$ and $m_2$, respectively. 
 The magnetic force $F_\mrm{mag}=F_\mrm{mag}(s,I)$ establishes the coupling between the mechanical and electromagnetic domains of the model. 
 This force is a function of the air gap~$s$ and the electric current~$I$ flowing through the electromagnet. 
 Consequently, it is essential to incorporate a model that captures the magnet's electromagnetic dynamics.

In this work, a detailed nonlinear model of the Transrapid’s levitation magnets is employed, which accounts for key phenomena such as magnetic saturation and eddy currents. 
The electromagnetic model for one half-magnet is described by an ordinary differential equation (ODE) governing the electric current, expressed as
\begin{equation}\label{eq:magnet_model}
     \dot{I} = \alpha(s,\dot{s},I) + \beta(s,I)U,
\end{equation}
where $s$ is the air gap, $\dot{s}$ the air gap velocity, $I$ the electric current, and $U$ the applied voltage. 
The nonlinear functions $\alpha(s,\dot{s},I)$ and $\beta(s,I)$ capture the complex electromagnetic behavior and are implemented using lookup tables, which are derived from a detailed equivalent magnetic circuit model~\cite{SchmidEtAl21}.
\begin{figure}[b]
     \includegraphics[width=\linewidth]{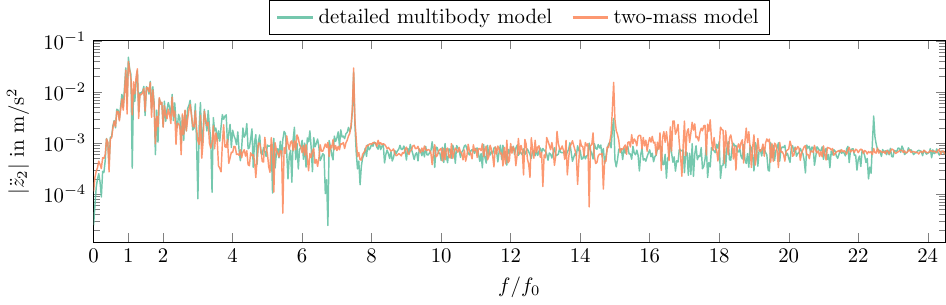}
     \centering
     \caption{Comparison of the acceleration spectrum between the middle section of the detailed multibody model and the two-mass model.}
     \label{fig:model_validation_spectrum}
 \end{figure}
The complete two-mass mechatronic model enables the prediction of both magnet and car body dynamics. 
To validate this simplified modeling approach, an exemplary ride over an uneven guideway with stochastic irregularities is simulated, and the resulting car body motion is analyzed.
For reference, a detailed multibody model of the vehicle as illustrated in Figure~\ref{fig:2D_model_schematic} is used. 
This detailed model includes 76 mechanical degrees of freedom and provides a comprehensive representation of the suspension system as well as the interaction between the magnets, the levitation chassis, and the car body~\cite{SchneiderEtAl21}.
The corresponding frequency spectrum of the car body movement is presented in Figure~\ref{fig:model_validation_spectrum}.
The simple two-mass model shows very similar behavior as the detailed model in the frequency range close to the first eigenfrequency~$f_0$.
This correspondence is expected, as the suspension parameters $c_k$ and $c_d$ are tuned to match the eigenfrequency~$f_0$ of the car body.
At higher frequencies, both models continue to show peaks corresponding to the primary excitation frequency of the guideway disturbance and its higher-order harmonics.
This confirms that the low frequent dynamics of the car body can be approximated with the much simpler two-mass model, motivating its utilization in the controller design.

\section{Controller Design}\label{sec:controller_design}
To achieve vertical vibration reduction and maintain stable levitation, a Nonlinear Model Predictive Control strategy is employed. 
The NMPC utilizes the previously introduced two-mass mechatronic model to predict the future behavior of the system and to determine an optimal control input by solving an Optimal Control Problem (OCP) over a finite prediction horizon of length~$T$. 
The core idea is to minimize a cost function that reflects the trade-off between air gap stability, ride comfort, and actuator effort, while respecting all relevant constraints. 
The control model defined by the coupled nonlinear system equations~\eqref{eq:two_mass_mechanics} and~\eqref{eq:magnet_model} is set up in state-space representation~$\dot{\bm{x}}=\bm{f}(\bm{x}, u)$.
The states and the input are defined with respect to their nominal equilibrium point and given by~$\bm{x} := \begin{bmatrix} \Delta s & \Delta z_2 & \dot{z}_1 & \dot{z}_2 & \Delta I \end{bmatrix}\T$ with $\Delta s = s-s_\mrm{nom}$, $\Delta I = I-I_\mrm{nom}$, and $u=U-U_\mrm{nom}$, respectively. 
The model output is chosen to~$\bm{y} := \begin{bmatrix} s & \Delta z_2 & \ddot{z}_1 & \ddot{z}_2 & I \end{bmatrix}\T$.
At each sampling instance~$t_k = k\delta$ with $k \in \mathbb{N}_0$ and sampling time $\delta>0$, the NMPC algorithm solves 
\begin{subequations}
     \begin{align}
          \min_{u(\cdot \at t_k)} \quad & \int_{t_k}^{t_k+T} \big(\bm{y}(t \at t_k)-\bm{y}_\mrm{ref}\big)\T \bm{Q} \big(\bm{y}(t \at t_k)-\bm{y}_\mrm{ref}\big) + R \big(u(t \at t_k)-u_\mrm{ref}\big)^2 \, \mrm{d}t \\
          \text{s.t.} \quad 
          & \bm{x}(t_k \at t_k) = \bm{x}(t_k) \,, \label{eq:OCP_initial_condition} \\ 
          & \dot{\bm{x}}(t) = \bm{f}\big(\bm{x}(t \at t_k), u(t \at t_k)\big), \qquad \forall t \in [t_k, t_k+T) \,, \label{eq:OCP_dynamic_contraint} \\ 
          & u(t) \in \mathcal{U}, \qquad \qquad \qquad \qquad \qquad \forall t \in [t_k, t_k+T) \,, \label{eq:OCP_input constraint} 
          \end{align}
          \label{eq:ocp}
\end{subequations}
where $\bm{Q} = \mrm{diag}(\begin{bmatrix} q_1 & q_2 & q_3 & q_4 & q_5\end{bmatrix})$ is a positive definite matrix, weighting the output error with respect to $\bm{y}_\mrm{ref} = \begin{bmatrix} s_\mrm{nom} & \Delta z_{2,\mrm{nom}} & 0 & 0 & I_\mrm{nom}\end{bmatrix}\T$. 
The weights $q_1$ and $q_5$ are primarily used to tune system stability, while $q_2$, $q_3$, and $q_4$ influence ride comfort.
Additionally, the weight $R>0$ penalizes the input with respect to $u_\mrm{ref} = U_\mrm{nom}$. 
Equations~\eqref{eq:OCP_initial_condition}-\eqref{eq:OCP_input constraint} denote the state's initial condition, the state dynamics as well as the input constraint, respectively.
The first value of the resulting optimal control input~$u_\mrm{MPC}=u^*(t_k \at t_k )$ is subsequently applied to the system for $t \in [t_k, t_k+\delta)$, until the next sampling time~$t_{k+1}$.
This process is repeated at each sampling instant, forming the basis of the receding horizon strategy that characterizes Model Predictive Control.
It is important to note that the proposed NMPC formulation implicitly assumes that the guideway disturbance stays constant over the prediction horizon, i.e., $\dot{d}_\mrm{gw}=0$.
Although this assumption limits the controller’s ability to anticipate the evolution of the disturbance, effective disturbance rejection is still achieved. 
No additional disturbance on the system is considered.
Moreover, it is assumed throughout this study that the full system state is directly measurable, thereby obviating the need for a state observer.

\subsection{Implementation}
The NMPC algorithm is implemented in MATLAB/Simulink, leveraging CasADi~\cite{AnderssonEtAl18} for symbolic gradient and Jacobian computation, and acados~\cite{VerschuerenEtAl21} for efficient real-time solution of the nonlinear optimal control problem.
The continuous-time OCP~\eqref{eq:ocp} is discretized over the prediction horizon of length $T$ into $N$ equidistant intervals using direct multiple shooting.
Both states and control inputs are treated as decision variables, with continuity constraints imposed to ensure trajectory consistency between intervals. 
The system dynamics is then integrated using an explicit Runge-Kutta scheme.
Direct multiple shooting exhibits superior numerical stability compared to gradient projection methods~\cite{EnglertEtAl19}, particularly for unstable systems with long prediction horizons.
It avoids the accumulation of integration errors by treating each shooting interval as an independent initial value problem rather than propagating states through the entire horizon sequentially.
The discretized OCP is subsequently solved via Sequential Quadratic Programming (SQP) using the high-performance interior point method (HPIPM) backend provided by acados.
For the proposed NMPC scheme, the sampling time is $\delta = 1\,\mrm{ms}$. 
The simulations are performed in Simulink on an AMD Ryzen 9 5950X 16-core processor (3.40 GHz, 64 GB DDR4).

\section{Simulation Results}\label{sec:results}
\begin{figure}[h]
     \centering
     \includegraphics{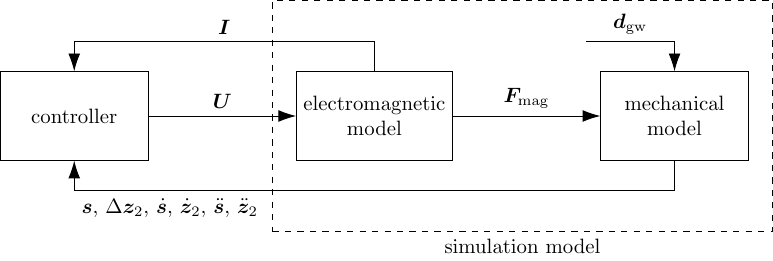}
     \caption{Signal flow between the controller and simulation model of the measured signals and the control variable.}
     \label{fig:signal_flow}
\end{figure}

The proposed NMPC algorithm is evaluated and compared to the standard approach with the single mass model through simulations based on a realistic ride scenario. 
These simulations are performed using two distinct simulation models, each incorporating a magnetic and mechanical component as depicted in Figure~\ref{fig:signal_flow}. While the magnetic models are taken from~\cite{SchmidEtAl21} and identical across both setups, the mechanical models differ substantially. 
The first simulation model employs the same two-mass system used during controller design, see Figure~\ref{fig:two_mass_model}.
A realistic guideway disturbance~$d_\mrm{gw}$ is applied to the mechanical model, which incorporates the bending of the guideway girders as well as stochastic irregularities~\cite{ZhengEtAl19}.

In contrast, the second simulation model provides a more comprehensive vehicle representation, utilizing a multibody model of the entire vehicle as illustrated in Figure~\ref{fig:2D_model_schematic}. 
This model is able to represent the vehicle's rigid body dynamics in the~$x$-$y$-plane with 76 degrees of freedom. 
In this configuration, 48 independent control units, each implementing one isolated control algorithm, operate simultaneously to maintain stable levitation of the whole vehicle.
Furthermore, the vehicle is riding on a flexible guideway of infinite length modeled by elastic beams~\cite{SchneiderEtAl21}.
The vehicle model is part of the software package TR.Mechatron~\cite{DignathEtAl25}, which additionally provides the necessary sensor and magnet models for testing the proposed NMPC algorithm.
In the subsequent section, three different controllers are compared, which are summarized in Table~\ref{tab:controller_configurations}.
The first controller, denoted as $\mrm{C}_\mrm{1M}$, is the baseline controller with the single mass model.
The other two controllers~$\mrm{C}_\mrm{2M}$ and $\mrm{C}_\mrm{2ML}$ utilize the two-mass model, where $\mrm{C}_\mrm{2ML}$ implements a longer prediction horizon.   
\begin{table}[h]
     \renewcommand{\arraystretch}{1.2}
     \centering
     \caption{The three controllers and their configurations used in the simulation study.}
     \begin{tabular}{cccc}
          controller name & controller model & prediction horizon & discretization intervals \\\hline 
          $\mrm{C}_\mrm{1M}$ & single mass & $50\,\mathrm{ms}$ & $50$ \\ 
          $\mrm{C}_\mrm{2M}$ & two-mass & $50\,\mathrm{ms}$ & $50$ \\ 
          $\mrm{C}_\mrm{2ML}$ & two-mass & $500\,\mathrm{ms}$ & $500$ \\ 
     \end{tabular}
     \label{tab:controller_configurations}
\end{table}

\subsection{Results with the two-mass simulation model} 
In this simulation scenario, the closed-loop performance is investigated at a vehicle speed of $v=600\,\mrm{km/h}$ with the two-mass simulation model.
The simulation is conducted for a duration of~$30\,\mrm{s}$.
The weights for $\mrm{C}_\mrm{2M}$ and $\mrm{C}_\mrm{2ML}$ are set to $\bm{Q} = \mrm{diag}(\begin{bmatrix} 10^2 & 1 & 1 & 1 & 10^5\end{bmatrix})$ and $R = 1$, in order to emphasize the reduction of car body accelerations while still stabilizing the overall system.
\begin{figure}[t]
     \centering
     \includegraphics[width=\linewidth]{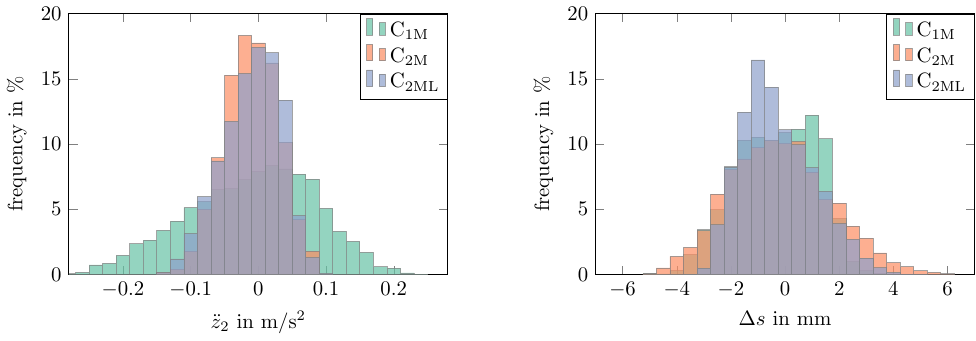} 
     \caption{Histograms of the car body acceleration (left) and air gap tracking error (right) for controllers~$\mrm{C}_{\mrm{1M}}$, $\mrm{C}_{\mrm{2M}}$, and $\mrm{C}_{\mrm{2ML}}$.}
     \label{fig:twoMass_comparison_hist}
\end{figure}

\begin{figure}
     \centering
     \includegraphics[width=\linewidth]{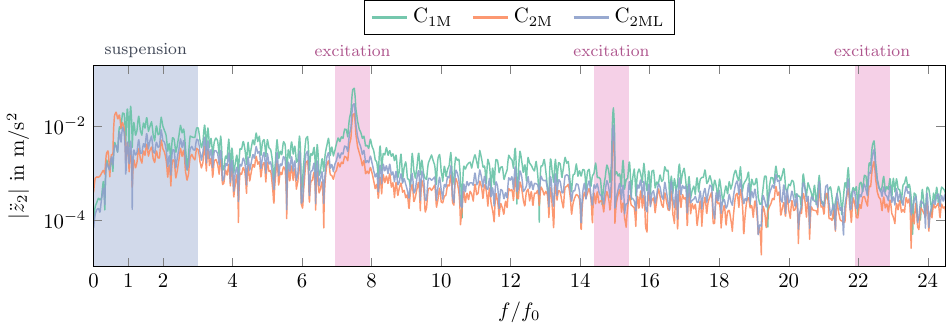} 
     \caption{Acceleration frequency spectrum of mass $m_2$ when using controllers~$\mrm{C}_{\mrm{1M}}$, $\mrm{C}_{\mrm{2M}}$, and $\mrm{C}_{\mrm{2ML}}$.}
     \label{fig:twoMass_comparison_spectrum}
\end{figure}

All three controllers result in a stable closed-loop with satisfactory performance.  
However, the acceleration histogram of the mass $m_2$, shown in Figure~\ref{fig:twoMass_comparison_hist}, reveals significant differences in the reached maximum accelerations.
While the controllers~$\mrm{C}_{\mrm{2M}}$ and~$\mrm{C}_{\mrm{2ML}}$ exhibit comparable performance with respect to maximum acceleration values, the standard controller~$\mrm{C}_{\mrm{1M}}$ imposes significantly higher accelerations on mass $m_2$, i.e., the car body. 
This discrepancy primarily arises from the standard controller’s limited mechanical model. 
It operates solely by influencing accelerations at magnet level, which introduces inherent limitations in its ability to control the car body dynamics effectively.
On the other hand, controller $\mrm{C}_{\mrm{1M}}$ achieves slightly smaller air gap deviations, as seen in Figure~\ref{fig:twoMass_comparison_hist}.
This can be attributed to the comparably high value for $q_5$ for controllers~$\mrm{C}_{\mrm{2M}}$~and~$\mrm{C}_{\mrm{2ML}}$ in this scenario, forcing the controllers to minimize the acceleration rather than the air gap deviation.
Further insights can be gained from Figure~\ref{fig:twoMass_comparison_spectrum}, where the acceleration frequency spectrum of mass $m_2$ is illustrated.
Evidently, controllers~$\mrm{C}_{\mrm{2M}}$ and~$\mrm{C}_{\mrm{2ML}}$ provide reduced accelerations across the evaluated frequency range.
Notably, around the suspension eigenfrequency~$f_0$, the long-horizon controller~$\mrm{C}_{\mrm{2ML}}$ performs particularly well, effectively attenuating low frequency vibrations of the car body mass~$m_2$.
In contrast, controller~$\mrm{C}_{\mrm{1M}}$ provides almost no additional damping in this frequency range.
The limited performance of controller~$\mrm{C}_{\mrm{2M}}$ at frequencies below~$f_0$ can be attributed to its short prediction horizon, since the slow car body dynamics appear almost constant over its short prediction horizon, preventing the controller from anticipating and compensating for them effectively in this scenario.
Overall, the results from this simulation underline the proposed NPMC controller's capability to effectively reduce the car body vibration in the low frequency spectrum.
Furthermore, some insights in the computation times of the NMPC algorithms are notable.
The average computation times for solving the OCP~\eqref{eq:ocp} are determined utilizing the Simulink profiler, resulting in $0.32\,\mrm{ms}$ for $\mrm{C}_{\mrm{1M}}$, $1.30\,\mrm{ms}$ for $\mrm{C}_{\mrm{2M}}$, and $24.0\,\mrm{ms}$ for $\mrm{C}_{\mrm{2ML}}$.

\subsection{Results with detailed multibody simulation model}
Now, the performance of the proposed NMPC algorithm is evaluated using the detailed multibody model of the Transrapid vehicle.
In this simulation scenario, the vehicle travels at a speed of $v=600\,\mrm{km/h}$ with a simulation duration of $30\,\mrm{s}$.
Here, the guideway excitation is caused by the elastic bending of the guideway girders. 
For a clearer comparison, the stochastic unevenness profile is not considered in this simulation, as the main excitation is caused by the bending guideway motion.
Again, controllers $\mrm{C}_{\mrm{1M}}$ and $\mrm{C}_{\mrm{2M}}$ are investigated with the weighting matrices set to $\bm{Q} = \mrm{diag}(\begin{bmatrix} 10^3 & 10^4 & 1 & 1 & 10^2\end{bmatrix})$ and $R = 1$.
The high computation time of controller $\mrm{C}_{\mrm{2ML}}$ due to the long prediction horizon makes its application in this scenario very difficult.
However, by increasing the control weight~$q_2$ for the car body position~$\Delta z_2$, similar performance may be achieved even with the shorter prediction horizon of controller $\mrm{C}_{\mrm{2M}}$.
\begin{figure}
     \centering
     \includegraphics[width=\linewidth]{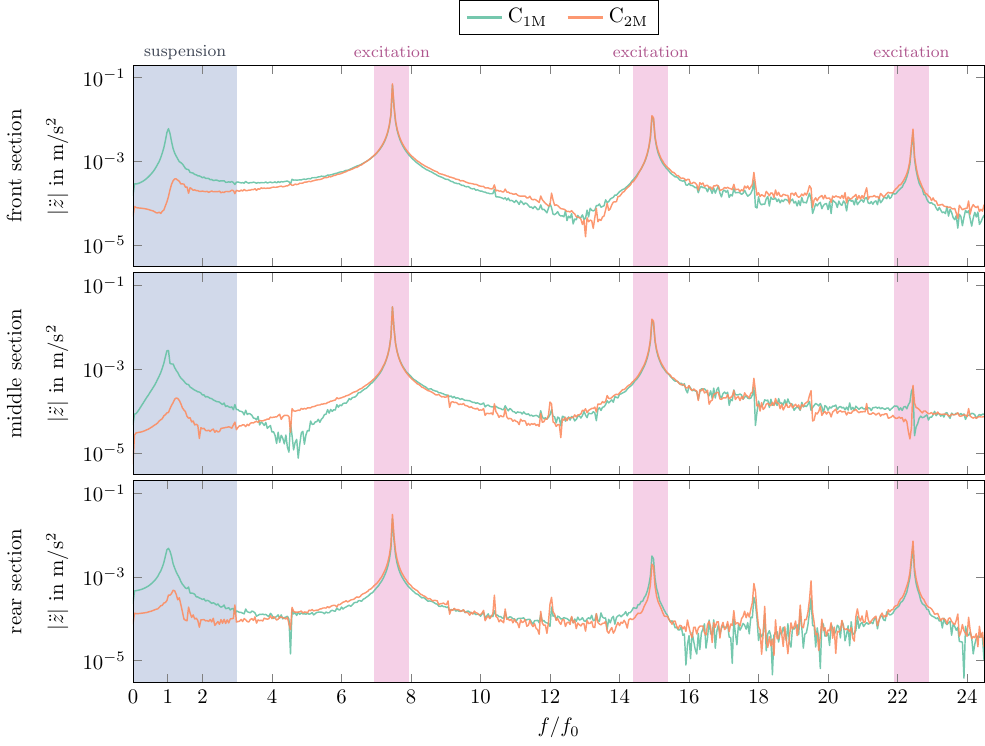}
     \caption{Frequency spectra of the car body acceleration for the front, middle, and rear section of the multibody simulation vehicle.}
     \label{fig:fhzg_vertikal_elast_carbody_spectrum}
\end{figure} 
\begin{figure}
     \centering
     \includegraphics{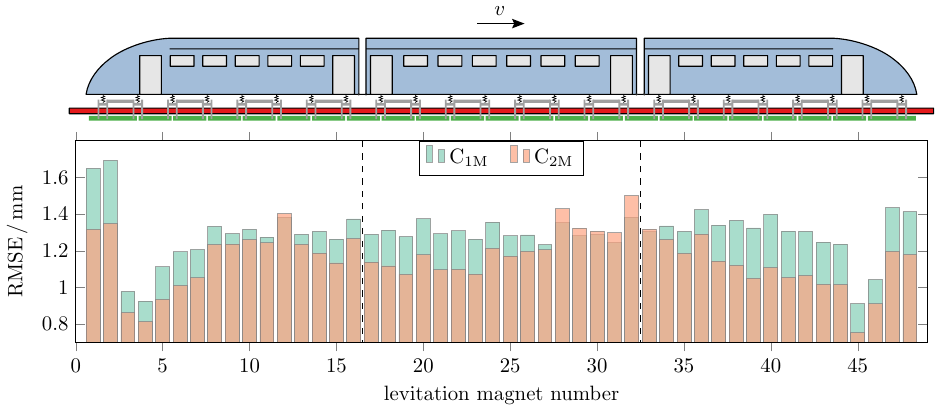}
     \caption{Root Mean Square Error (RMSE) of the air gap tracking error for all levitation magnets of the multibody simulation vehicle.}
     \label{fig:fhzg_vertikal_elast_RMSE}
\end{figure}
Figure~\ref{fig:fhzg_vertikal_elast_carbody_spectrum} shows the frequency spectrum of the car body acceleration for the front, middle, and rear section of the vehicle.
The results show that the NMPC controller $\mrm{C}_{\mrm{2M}}$ significantly reduces the low frequency car body vibrations across all vehicle sections compared to the standard controller~$\mrm{C}_{\mrm{1M}}$ in the relevant range of~$1-5\,\mrm{Hz}$.
However, at higher excitation frequencies caused by the guideway disturbance, no improvement is observed.
This is expected, as the NMPC controller is not able to predict the guideway disturbance, but only the low frequency car body movements.
This confirms that even with the distributed control architecture of the Transrapid, the NMPC controller is able to effectively reduce the low frequency car body vibrations.
At the same time, the air gap tracking error is also improved, as seen in~Figure~\ref{fig:fhzg_vertikal_elast_RMSE}.
The Root Mean Square Error (RMSE) of the air gap tracking error is smaller for almost all levitation magnet when employing controller $\mrm{C}_{\mrm{2M}}$. 
The improved air gap tracking can be attributed to the different weighting matrices chosen in this simulation scenario, which emphasizes the importance of finding the right balance between air gap stability and ride comfort. 
However, with the proposed two-mass control model, this trade-off can be adjusted more flexibly compared to the standard controller due to the additional weighting factors available in the cost function.
Overall, the proposed NMPC strategy demonstrates promising potential to enhance both stability and passenger comfort in high-speed Maglev systems.

\section{Conclusion}
This work presents a novel Nonlinear Model Predictive Control (NMPC) strategy for high-speed Maglev vehicles that explicitly incorporates mechanical suspension dynamics into the control model. 
Unlike conventional approaches, the proposed method enables predictive vibration mitigation by modeling both electromagnetic forces and suspension behavior. 
This integrated approach significantly improves passenger comfort and ride quality by reducing low frequency vertical oscillations of the car body. 
Moreover, it allows for a more direct tuning of the trade-off between air gap tracking and ride comfort compared to previous methods. 
The simulation results, validated against both the simplified and the detailed multibody model, demonstrate better performance compared to the conventional single-mass approach, especially in suppressing low frequency vibrations that significantly impact passenger comfort.
While the proposed approach shows promising results, some limitations should be noted. 
The computational complexity increases substantially with longer prediction horizons, potentially limiting real-time implementation on the available embedded hardware. 
Additionally, the current implementation assumes perfect state measurement and constant guideway disturbance over the prediction horizon, which may not be realistic in practice. 
Furthermore, the trade-off between air gap stability and ride comfort requires careful balancing through weighting factors in the cost function.
The demonstrated improvements in vibration suppression, particularly in the critical low-frequency range, make this approach a promising solution for future high-speed Maglev applications.

\vspace{\baselineskip}
\bibliographystyle{pamm}
\providecommand{\WileyBibTextsc}{}
\let\textsc\WileyBibTextsc
\providecommand{\othercit}{}
\providecommand{\jr}[1]{#1}
\providecommand{\etal}{~et~al.}

\end{document}